\documentclass[english,english,12pt,onecolumn, draftcls]{IEEEtran}
\usepackage[T1]{fontenc}
\usepackage[latin9]{inputenc}
\usepackage{units}
\usepackage{amsthm}
\usepackage{amsmath}
\usepackage{graphicx}
\usepackage{esint}

\makeatletter
\theoremstyle{plain}
\newtheorem{thm}{\protect\theoremname}
\theoremstyle{definition}
\newtheorem{defn}[thm]{\protect\definitionname}
\theoremstyle{plain}
\newtheorem{prop}[thm]{\protect\propositionname}
\theoremstyle{remark}
\newtheorem{rem}[thm]{\protect\remarkname}

\ifCLASSINFOpdf
\else
\fi
\makeatother

\usepackage{babel}
\providecommand{\definitionname}{Definition}
\providecommand{\propositionname}{Proposition}
\providecommand{\remarkname}{Remark}
\providecommand{\theoremname}{Theorem}

\begin{document}

\title{On Channels with Action-Dependent States}

\author{Behzad~Ahmadi and Osvaldo Simeone 
\thanks{B. Ahmadi and O. Simeone are with the CWCSPR, New Jersey Institute
of Technology, Newark, NJ 07102 USA (e-mail: \{behzad.ahmadi,osvaldo.simeone\}@njit.edu). %
}
}
\maketitle
\begin{abstract}
Action-dependent channels model scenarios in which transmission takes
place in two successive phases. In the first phase, the encoder selects
an \textquotedbl{}action\textquotedbl{} sequence, with the twofold
aim of conveying information to the receiver and of affecting in a
desired way the state of the channel to be used in the second phase.
In the second phase, communication takes place in the presence the
mentioned action-dependent state. In this work, two extensions of
the original action-dependent channel are studied. In the first, the
decoder is interested in estimating not only the message, but also
the state sequence within an average per-letter distortion. Under
the constraint of common reconstruction (i.e., the decoder's estimate
of the state must be recoverable also at the encoder) and assuming
non-causal state knowledge at the encoder in the second phase, we
obtain a single-letter characterization of the achievable rate-distortion-cost
trade-off. In the second extension, we study an action-dependent degraded
broadcast channel. Under the assumption that the encoder knows the
state sequence causally in the second phase, the capacity-cost region
is identified. Various examples, including Gaussian channels and a
model with a \textquotedbl{}probing\textquotedbl{} encoder, are also
provided to show the advantage of a proper joint design of the two
communication phases. \end{abstract}
\IEEEpeerreviewmaketitle

\section{Introduction}

In \cite{Weissman}, the framework of action-dependent channels was
introduced as a means to model scenarios in which transmission takes
place in two successive phases. In the first phase, the encoder selects
an \textquotedbl{}action\textquotedbl{} sequence, with the twofold
aim of conveying information to the receiver and of affecting in a
desired way the state of the channel to be used in the second phase.
In the second phase, communication takes place in the presence the
mentioned action-dependent state. With a cost constraint on the actions
in the first phase and on the channel input in the second phase, reference
\cite{Weissman} derived the capacity-cost-trade-off under the assumption
that the channel state is available either causally or non-causally
at the encoder in the second phase. 

A number of applications and extensions of the results in \cite{Weissman}
have been reported since then. In \cite{probing}, the result in \cite{Weissman}
is leveraged to study a model in which encoder and decoder can \textquotedbl{}probe\textquotedbl{}
the channel state to obtain partial state information during the first
communication phase. In \cite{KTH}, unlike \cite{Weissman} the decoder
is required to decode both the transmitted message and channel input
reliably. Finally, in \cite{Chiru}, the decoder is interested in
estimating not only the message but also the state sequence, and the
latter is available strictly causally at the encoder in the second
transmission phase. 

In this paper, two further extensions of the original action-dependent
channel are studied. In the first, similar to \cite{Chiru}, the decoder
is interested in estimating not only the message but also the state
sequence within given average per-letter distortion constraints (see
Fig. 1). Unlike \cite{Chiru}, we assume \emph{non-causal} state knowledge
in the second phase, and, under the constraint of common reconstruction
(CR) (i.e., the decoder's estimate of the state must be recoverable
also at the encoder with high probability \cite{Steinberg}), we obtain
a single-letter characterization of the achievable rate-distortion-cost
trade-off. We remark that, for conventional state-dependent states
without actions, the problem of joint estimation of message and state
with non-causal state information at the encoder without the CR constraint
is open (see, e.g., \cite{KSC}), while with the CR constraint the
problem has been solved in \cite{Steinberg}. In the second extension,
illustrated in Fig. \ref{fig:fig2}, we study an action-dependent
degraded broadcast channel. Under the assumption that the encoder
knows the state sequence causally in the second phase, the capacity-cost
region is identified.%
\footnote{After submitting \cite{Ahmadi}, we have been informed of the reference
\cite{Steinberg-ISIT'12}, where the problem illustrated in Fig. \ref{fig:fig2}
has also been solved.%
} The corresponding result for action-independent states was derived
in \cite{Steinberg_IT05} (see also \cite{Kim}), while we recall
that with non-causal state information the problem is open (see \cite{Steinberg_Shamai}).
Various examples, including Gaussian channels and a model with a \textquotedbl{}probing\textquotedbl{}
encoder, are also provided throughout to show the advantage of a proper
joint design of the two communication phases. 
\begin{figure}[h!]
\centering \includegraphics[scale=0.65]{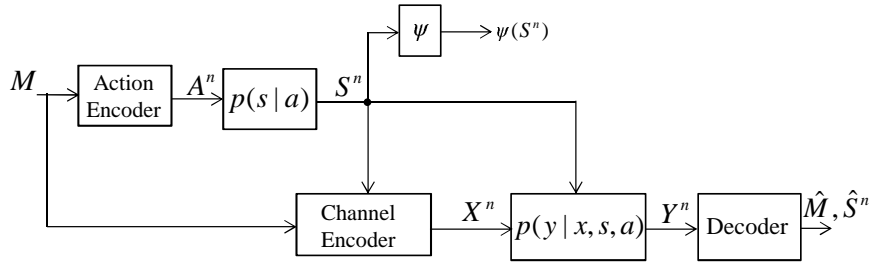} \caption{Channel with action-dependent state in which the decoder estimates
both message and state, and there is a a common reconstruction (CR)
constraint on the state reconstruction. The state is known non-causally
at the channel encoder.}

\label{fig:fig1}
\end{figure}

\section{Transmission of Data and Action-Dependent State with Common Reconstruction
Constraint}

In this section, we study the setting illustrated in Fig.\ref{fig:fig1}
of a channel with action-dependent state in which the decoder estimates
both message and state. We first detail the system model in Sec. \ref{sub:System-model_act}.
Next, the characterization of the trade-off between the achievable
data rate and state reconstruction distortion is derived in Sec. \ref{sub:Capacity_Fun}.
Finally, a Gaussian example is given in Sec. \ref{sub:Gaussian-Example}.


\begin{figure}[h!]
\centering \includegraphics[scale=0.65]{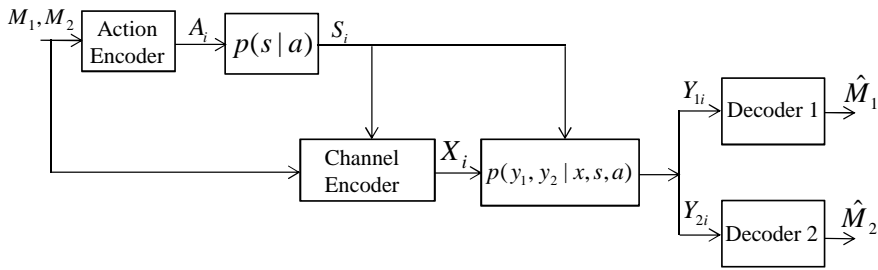} \caption{Broadcast channel with action-dependent states known causally to the
encoder (i.e., the $i$th transmitted symbol $X_{i}$ is a function
of messages $M_{1}$, $M_{2}$ and the state symbols up to time $i$,
$S^{i}$).}

\label{fig:fig2}
\end{figure}

\subsection{System Model\label{sub:System-model_act}}

In this section the system model is detailed. The system is defined
by the probability mass functions (pmfs) $p(x)$, $p(y|x,s,a)$, $p(s|a)$
and discrete alphabets $\mathcal{X},\mathcal{A\mbox{, }S\mbox{,\ensuremath{\mathcal{\mbox{ }\hat{S}}}, and }Y}$
as follows. Given the message $M$, selected randomly from the set
$\mathcal{M}=[1,2^{nR}]$, an action sequence $A^{n}\in\mathcal{A}^{n}$
is selected. As a result of this selection, the state sequence $S^{n}\in\mathcal{S}^{n}$
is generated as the output of a memoryless channel $p(s|a)$ so that
we have $p(s^{n}|a^{n})=\prod_{i=1}^{n}p(s_{i}|a_{i})$ for an action
sequence $A^{n}=a^{n}$.  The input sequence $X^{n}\in\mathcal{X}^{n}$
is selected based on both message $M$ and state sequence $S^{n}$.
The action sequence $A^{n}$ and the input $X^{n}$ have to satisfy
an average cost constraint defined by a function $\gamma:\mathcal{A}\times\mathcal{X}\rightarrow[0,\infty)$,
so that the cost for the input sequences $a^{n}$ and $x^{n}$ is
given by $\gamma(a^{n},x^{n})=\frac{1}{n}\sum_{i=1}^{n}\gamma(a_{i},x_{i}).$
Given $X^{n}=x^{n}$, $S^{n}=s^{n}$ and $A^{n}=a^{n}$, the received
signal is distributed as $p(y^{n}|x^{n},s^{n},a^{n})=\prod_{i=1}^{n}p(y_{i}|x_{i},s_{i},a_{i})$.
The decoder, based on the received signal $Y^{n}$, estimates the
message $M$ and the sequences $S^{n}\in\mathcal{S}^{n}$. The estimate
$\hat{S}^{n}\in\mathcal{\hat{S}}^{n}$ is constrained to satisfy a
distortion criterion defined by a per-symbol distortion metric $d(s,\hat{s}):\mathcal{S}\times\mathcal{\hat{S}}\rightarrow[0,D_{max}]$
with $0<D_{max}<\infty$. Based on the given distortion metric, the
overall distortion for the estimated state sequences $\hat{s}^{n}$
is defined as $d^{n}(s^{n},\hat{s}^{n})=\frac{1}{n}\sum_{i=1}^{n}d(s_{i},\hat{s}_{i})\mbox{. }$The
reconstructions $\hat{S}^{n}$ is also required to satisfy the CR
constraint, which imposes that the state estimate be also reproducible
at the encoder with high probability, as formalized below.
\begin{defn}
\label{def_pp}An $(n,R,D,\Gamma,\epsilon)$ code for the model in
Fig. \ref{fig:fig1} consists of an action encoder
\begin{align}
g_{1}\textrm{: }\mathcal{M}\rightarrow\mathcal{A}^{n},\label{eq:action_enc}
\end{align}
which maps message $M$ into an action sequence $A^{n}$; a channel
encoder
\begin{align}
g_{2}\textrm{: }\mathcal{M}\times\mathcal{S}^{n}\rightarrow\mathcal{X}^{n},\label{eq:enc}
\end{align}
which maps message $\mbox{\ensuremath{M}}$ and the state sequence
$S^{n}$ into the sequence $X^{n}$; two decoding functions,
\begin{align}
h_{1}\textrm{: }\mathcal{Y}^{n}\rightarrow\mathcal{M},\\
\textrm{and }h_{2}\textrm{: }\mathcal{Y}^{n}\rightarrow\hat{\mathcal{S}}^{n},
\end{align}
which map the sequence $Y_{1}^{n}$ into the estimated message $\hat{M}$
and into the estimated sequence $\hat{S}^{n}$, respectively; and
a reconstruction function
\begin{align}
\psi\textrm{: }\mathcal{S}^{n}\rightarrow\hat{\mathcal{S}}^{n},
\end{align}
which maps the state sequence into the estimated state sequence at
the encoder; such that the probability of error in decoding the message
$M$ is small
\begin{eqnarray}
\textrm{Pr}[\hat{M}\neq M] & \leq & \epsilon,\label{eq:small_err}
\end{eqnarray}
the distortion and cost constraints are satisfied, i.e., 
\begin{eqnarray}
\frac{1}{n}\sum_{i=1}^{n}\textrm{E}\left[d(S_{i},\textrm{\ensuremath{h_{2i}}}(Y^{n}))\right] & \leq & D+\epsilon\label{eq:dist_const}\\
\textrm{and }\frac{1}{n}\sum_{i=1}^{n}\textrm{E}\left[\gamma(A_{i},X_{i})\right] & \leq & \Gamma+\epsilon,\label{eq:cost_const}
\end{eqnarray}
where $\textrm{\ensuremath{h_{2i}}}(Y^{n})\in\hat{\mathcal{S}}$ is
the $i$th symbol of the sequence $\textrm{\ensuremath{h_{2}}}(Y^{n})$,
and the CR requirement is verified, namely, 
\begin{eqnarray}
\textrm{Pr}\left[\psi(S^{n})\neq h_{2}(Y^{n})\right] & \leq & \epsilon.\label{eq:CK_req}
\end{eqnarray}
We note that, given the definition above, the pmf of the random variables
$(M,A^{n},S^{n},X^{n},Y^{n})$ factorizes as
\begin{eqnarray}
p(m,a^{n},s^{n},x^{n},y^{n}) & = & \frac{1}{2^{nR}}\delta[a^{n}-\textrm{\ensuremath{g_{1}}}(m)]\left\{ \prod_{i=1}^{n}p(s_{i}|a_{i})\right\} \delta[x^{n}-\textrm{\ensuremath{g_{2}}}(m,s^{n})]\nonumber \\
 &  & \cdot\left\{ \prod_{i=1}^{n}p(y_{i}|x_{i},s_{i},a_{i})\right\} ,
\end{eqnarray}
where $\delta[\cdot]$ is the Kronecker delta function (i.e., $\delta[x]=1$
if $x=0$ and $\delta[x]=0$ otherwise) and the arguments of the pmf
range in the alphabets of the corresponding random variables. 

Given a cost-distortion pair $(D,\Gamma)$, a rate $R$ is said to
be achievable if, for any $\epsilon>0$ and sufficiently large $n$,
there a exists a $(n,R,D,\Gamma,\epsilon)$ code. We are interested
in characterizing the capacity-distortion-cost trade-off function
 $C(D,\Gamma)=$inf$\{R:$ the triple $(R,D,\Gamma)$ is achievable\}.
\end{defn}

\subsection{Capacity-Distortion-Cost Function\label{sub:Capacity_Fun}}

In this section, a single-letter characterization of the capacity-distortion-cost
function is derived.
\begin{prop}
\label{prop:act_stat}The capacity-distortion-cost function for the
system in Fig. \ref{fig:fig1} is given by 
\begin{eqnarray}
C(D,\Gamma) & = & \textrm{\ensuremath{\max}\ }I(U;Y)-I(U;S|A)\label{eq:C}
\end{eqnarray}
where the mutual informations are evaluated with respect to the joint
pmf
\begin{align}
p(a,u,s,x,y)=p(a)p(s|a)p(u|s,a)p(x|u,s)p(y|x,s,a) & ,\label{eq:joint}
\end{align}
and minimization is done with respect to the pmfs $p(a)$, $p(u|s,a)$
and $p(x|u,s)$ under the constraint that there exists a deterministic
function $\phi:\mathcal{U}\rightarrow\mathcal{\hat{S}}$ such that
the inequalities\begin{subequations}\label{eqn: constr} \textup{
\begin{eqnarray}
\mathrm{E}[d(S,\phi(U))] & \leq & D\label{eq:dist_bound}\\
\mathrm{\textrm{and }E}[\gamma(A,X)] & \leq & \Gamma\label{eq:cost_bound}
\end{eqnarray}
}\end{subequations}are satisfied. Finally, $U$ is an auxiliary random
variable whose alphabet cardinality can be bounded as $|\mathcal{U}|\leq|\mathcal{A}||\mathcal{S}||\mathcal{X}|+2$.\end{prop}
\begin{rem}
If we let $D\geq D_{max}$, the result above recovers Theorem 1 of
\cite{Weissman}. If instead we have $p(s|a)=p(s)$ so that the channel
is not action-dependent, we recover Theorem 1 in \cite{Steinberg}. 

The proof of achievability follows using the same arguments as in
\cite{Weissman} with the difference that here $U$ is also used to
estimate the state $S$ via a function $\phi(U)$. The proof of the
converse can be found in Appendix A.
\end{rem}

\subsection{A Gaussian Example\label{sub:Gaussian-Example}}

\begin{figure}[h!]
\centering \includegraphics[bb=45bp 175bp 559bp 593bp,clip,scale=0.65]{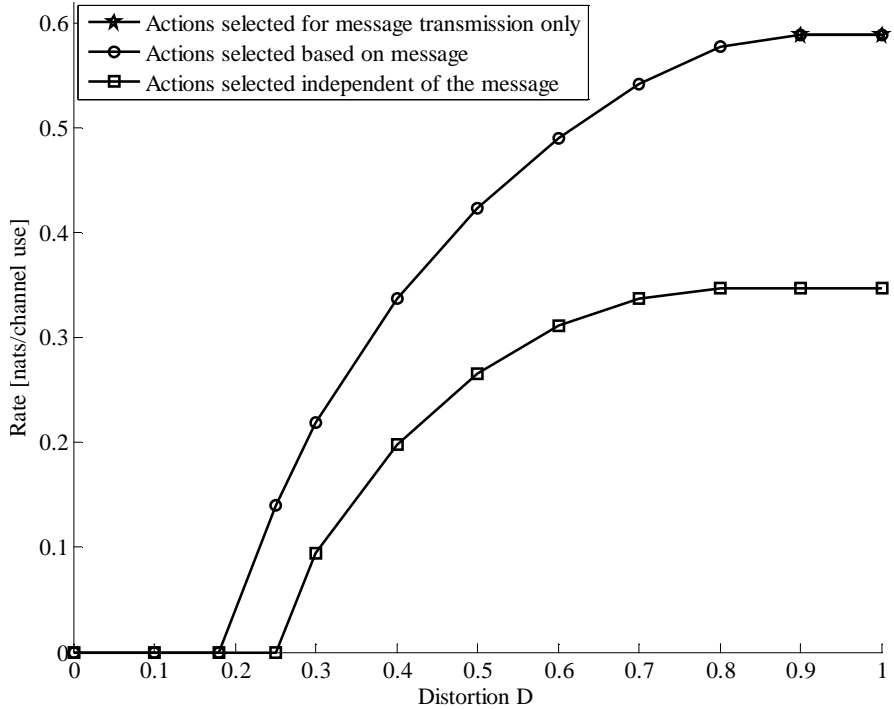}
\caption{Achievable rates (constrained to a Gaussian joint distribution, see
(\ref{eqn: joint_guass})) for the Gaussian model (\ref{eq:act_chan_Gauss})-(\ref{eq:Gauss_dec})
versus distortion $D$ for $P_{A}=P_{X}=\sigma_{W}^{2}=\sigma_{Z}^{2}=1$. }

\label{fig:plot}
\end{figure}

In this section, we consider a continous-alphabet version of the model
of Fig. \ref{fig:fig1} in which the actions and the channel input
are subject to the cost constraints $1/n\sum_{i=1}^{n}\textrm{E}[A{}^{2}]\leq P_{A}$
and $\frac{1}{n}\sum_{i=1}^{n}\textrm{E}\left[X_{i}^{2}\right]\leq P_{X},$
respectively; the action channel is given by
\begin{eqnarray}
S & = & A+W,\label{eq:act_chan_Gauss}
\end{eqnarray}
where $W\sim\mathcal{N}(0,\sigma_{W}^{2})$ and the transmission channel
is given by
\begin{eqnarray}
Y & = & X+S+Z,\label{eq:Gauss_dec}
\end{eqnarray}
where $Z\sim\mathcal{N}(0,\sigma_{Z}^{2})$ is independent of $W$.
We evaluate the rate $I(U;Y)-I(U;S|A)$ in (\ref{eq:C}) by assuming
the variables $(A,S,U,X,Y)$ to be jointly Gaussian without claiming
the optimality of this choice. Specifically, similar to \cite[Sec. VI]{Weissman},
we choose $A\sim\mathcal{N}(0,P_{A}),$\begin{subequations}\label{eqn: joint_guass}
\begin{eqnarray}
X & = & \alpha A+\gamma W+G\\
\textrm{and }U & = & \delta X+A+\beta W,\label{eq:U}
\end{eqnarray}
\end{subequations}with $G\sim\mathcal{N}(0,P_{X}-(\alpha^{2}P_{A}+\gamma^{2}\sigma_{W}^{2}))$,
where we enforce the constraint $P_{X}\geq(\alpha^{2}P_{A}+\gamma^{2}\sigma_{W}^{2}),$
and the variables $(A,W,G,Z)$ are all independent of each other.
We evaluate then the rate $I(U;Y)-I(U;S|A)$ as in \cite{Weissman},
with the difference that we have the additional constraint (\ref{eq:dist_const})
on the state estimate $\hat{S}$. Assuming the quadratic distortion
metric $d(s,\hat{s})=(s-\hat{s})^{2}$, we choose $\hat{S}$ to be
the MMSE estimate of $S$ given $U$ and $A$.%
\footnote{Note that $U$ in the characterization of Proposition 2 can be always
redefined to include also $A$ without loss of performance, and hence
$\hat{S}$ can be made to be a function of $U$ and $A$.%
} This leads to the constraint
\begin{eqnarray}
D\geq E[(S-\hat{S})^{2}]=var(S|U,A) & = & \sigma_{W}^{2}-\frac{(E[W(U-A)])^{2}}{E[(U-A)^{2}]},\label{eq:const_Gauss}
\end{eqnarray}
where $E[W(U-A)]=(\delta\gamma+\beta)\sigma_{W}^{2}$ $\textrm{and }E[(U-A)^{2}]=\delta{}^{2}P_{G}+(\delta\gamma+\beta)^{2}\sigma_{W}^{2}.$
The rate $I(U;Y)-I(U;S|A)$ optimized over parameters $(\alpha,\beta,\delta,\gamma)$
under the constraint (\ref{eq:const_Gauss}) for different values
of the distortion $D$ for $P_{A}=P_{X}=\sigma_{W}^{2}=\sigma_{Z}^{2}=1$
in Fig. (\ref{fig:plot}). Moreover for reference, Fig. \ref{fig:plot}
shows also the rate achievable if distribution (\ref{eq:joint}) is
designed to be optimal for message transmission only as in \cite[eq. (95)]{Weissman},
and the rate achievable, if $A$ is selected to be independent of
the message, namely, $\textrm{\ensuremath{\max}\ }I(U;Y|A)-I(U;S|A)$,
where the mutual information terms are evaluated with respect to the
joint Gaussian distribution given above in (\ref{eqn: joint_guass})
under the constraint (\ref{eq:const_Gauss}). The performance gains
attainable by designing the transmission strategy jointly in the two
phases and by accounting for the constraint (\ref{eq:const_Gauss})
are apparent.

\section{Degraded Broadcast Channels with Action-Dependent States}

In this section, we study the problem illustrated in Fig. 2 of a broadcast
channel with action-dependent states known causally to the encoder.
We first detail the system model in Sec. \ref{sub:System-model_act_DBC}.
Next, the characterization of the capacity region for physically degraded
broadcast channels is given in Sec. \ref{sub:Capacity_reg_DBC}. In
Sec. \ref{sub:Probing}, we study the special case of a broadcast
channel with a probing encoder in the sense of \cite{probing}.

\subsection{System Model\label{sub:System-model_act_DBC}}

In this section the system model is detailed. The system is defined
by the pmfs $p(x)$, $p(y_{1},y_{2}|x,$ $s,a)$, $p(s|a)$ and discrete
alphabets $\mathcal{X},\mathcal{\textrm{ }A\mbox{,}\textrm{ }S\mbox{, and }Y}$
as follows. Given the messages $M_{1}$ and $M_{2}$, selected randomly
from the sets $\mathcal{M}_{1}=[1,2^{nR_{1}}]$ and $\mathcal{M}_{2}=[1,2^{nR_{2}}]$,
respectively, an action sequence $A^{n}\in\mathcal{A}^{n}$ is selected.
As a result of this selection, the state sequence $S^{n}\in\mathcal{S}^{n}$
is generated as in the previous section.  The action sequence $A^{n}$
and the input $X^{n}$ have to satisfy the average cost constraint
(\ref{eq:cost_const}). Given the transmitted signal $X^{n}=x^{n}$,
the state sequence $S^{n}=s^{n}$, and the action sequence $A^{n}=a^{n}$,
the received signals are distributed as $p(y_{1}^{n},y_{2}^{n}|x^{n},s^{n},a^{n})=\prod_{i=1}^{n}p(y_{1i},y_{2i}|x_{i},s_{i},a_{i})$.
The decoders, based on the received signals $Y_{1}^{n}$ and $Y_{2}^{n}$
, estimate the messages $M_{1}$ and $M_{2},$ respectively.
\begin{defn}
\label{def_DBC}An $(n,R_{1},R_{2},\Gamma,\epsilon)$ code for the
model in Fig. \ref{fig:fig2} consists of an action encoder
\begin{align}
g_{1}\textrm{: }\mathcal{M}_{1}\times\mathcal{M}_{2}\rightarrow\mathcal{A}^{n},\label{eq:action_enc_DBC}
\end{align}
which maps messages $M_{1}$ and $M_{2}$ into an action sequence
$A^{n}$; a sequence of channel encoders
\begin{align}
g_{2i}\textrm{: }\mathcal{M}_{1}\times\mathcal{M}_{2}\times\mathcal{S}^{i}\rightarrow\mathcal{X},\label{eq:enc_DBC}
\end{align}
for $i\in[1,n]$ which map messages $M_{1}$ and $M_{2}$ and the
first $i$ samples of the state sequence $S^{i}$ into the $i$th
symbol $X_{i}$; two decoding functions,
\begin{align}
h_{1}\textrm{: }\mathcal{Y}_{1}^{n}\rightarrow\mathcal{M}_{1},\\
\textrm{and }h_{2}\textrm{: }\mathcal{Y}_{2}^{n}\rightarrow\mathcal{M}_{2},
\end{align}
which map the received sequences $Y_{1}^{n}$ and $Y_{2}^{n}$ into
the estimated messages $\hat{M}_{1}$ and $\hat{M}_{2}$, respectively;
such that the probability of error in decoding the messages $M_{1}$
and $M_{2}$ is small
\begin{eqnarray}
\textrm{Pr}[\hat{M}_{j}\neq M_{j}] & \leq & \epsilon\textrm{ for }j=1,2,\label{eq:small_err_DBC}
\end{eqnarray}
and the cost constraint (\ref{eq:cost_const}) is satisfied.
\end{defn}
We note that, given the definitions above, the distribution of the
random variables $(M_{1},M_{2},A^{n},$ $S^{n},X^{n},Y_{1}^{n},Y_{2}^{n})$
factorizes as
\begin{eqnarray}
p(m_{1},m_{2},a^{n},s^{n},x^{n},y_{1}^{n},y_{2}^{n}) & = & \frac{1}{2^{n(R_{1}+R_{2})}}\delta[a^{n}-\textrm{\ensuremath{g_{1}}}(m_{1},m_{2})]\left\{ \prod_{i=1}^{n}p(s_{i}|a_{i})\right\} \nonumber \\
 &  & \cdot\left\{ \prod_{i=1}^{n}\delta[x_{i}-\textrm{\ensuremath{g_{2i}}}(m_{1},m_{2},s^{i})]p(y_{1i},y_{2i}|x_{i},s_{i},a_{i})\right\} \negmedspace,\label{eq:code_dist_DBC}
\end{eqnarray}
where the arguments of the pmf range in the alphabets of the corresponding
random variables. 

Given a cost $\Gamma$, a rate pair $(R_{1},R_{2})$ is said to be
achievable if, for any $\epsilon>0$ and sufficiently large $n$,
there a exists a $(n,R_{1},R_{2},\Gamma,\epsilon)$ code. The capacity
region $\mathcal{C}(\Gamma)$ is defined as the closure of all rate
pairs $(R_{1},R_{2})$ that are achievable given the cost $\Gamma$.

\subsection{Capacity-Cost Region\label{sub:Capacity_reg_DBC}}

In this section, a single-letter characterization of the capacity
region is derived for the special case in which the channel is physically
degraded in the sense that we have the condition
\begin{align}
p(y_{1},y_{2}|x,s,a)=p(y_{1}|x,s,a)p(y_{2}|y_{1}) & ,\label{eq:degradedness}
\end{align}
or equivalently we have that the Markov chain $(X_{i},S_{i},A_{i})-Y_{1i}-Y_{2i}$
holds for all $i\in[1,n]$.
\begin{prop}
\label{prop:act_stat_DBC}The capacity region of the system in Fig.
\ref{fig:fig2} under the degradedness condition (\ref{eq:degradedness})
is given by the union of the rate pairs $(R_{1},R_{2})$ satisfying
\begin{subequations}\label{eqn: act_stat_DBC}
\begin{eqnarray}
R_{1} & \leq & I(U_{1};Y_{1}|U_{2})\label{eq:R1_BC}\\
\textrm{and }R_{2} & \leq & I(U_{2};Y_{2}),\label{eq:R2_BC}
\end{eqnarray}
\end{subequations}where the mutual informations are evaluated with
respect to the joint pmf
\begin{align}
p(a,u_{1},u_{2},s,x,y_{1},y_{2})=p(u_{1},u_{2})\delta[a-\textrm{\ensuremath{f_{a}}}(u_{1},u_{2})]p(s|a)\delta[x-\textrm{\ensuremath{f_{x}}}(u_{1},u_{2},s)]p(y_{1}|x,s,a)p(y_{2}|y_{1}),\label{eq:joint_DBC}
\end{align}
for some pmfs $p(u_{1},u_{2})$ and deterministic functions $\textrm{\ensuremath{f_{a}}}\textrm{: }\mathcal{U}_{1}\times\mathcal{U}_{2}\rightarrow\mathcal{A}$
and \textup{$\textrm{\ensuremath{f_{x}}}\textrm{: }\mathcal{U}_{1}\times\mathcal{U}_{2}\times\mathcal{S}\rightarrow\mathcal{X}$}
such that the inequality \textup{$\mathrm{E}[\gamma(A,X)]\leq\Gamma$}
is satisfied. Auxiliary random variables $U_{1}$ and $U_{2}$ have
finite alphabets.
\end{prop}
The proof of achievability can be sketched as follows. The codewords
$u_{2}^{n}(m_{2})$, encoding message $m_{2}\in[1,2^{nR_{2}}]$, are
generated independently and i.i.d. according to the pmf $p(u_{2})$.
Then, superimposed on each codeword $u_{2}^{n}(m_{2})$, $2^{nR_{1}}$
codewords $u_{1}^{n}(m_{1},m_{2})$ are generated independently according
to the distribution $\prod_{i=1}^{n}p(u_{1i}|u_{2i}(m_{2}))$. To
encode messages $(M_{1},M_{2})$, the action sequence $A^{n}$ is
obtained as a deterministic function of $u_{1i}(M_{1},M_{2})$ and
$u_{2i}(M_{2})$ such that $A_{i}=\textrm{\ensuremath{f_{a}}}(u_{1i}(M_{1},M_{2}),u_{2i}(M_{2}))$
for all $i\in[1,n]$. The transmitted symbol $X_{i}$ is obtained
instead as a function of $u_{1i}(M_{1},M_{2})$, $u_{2i}(M_{2})$,
and of the $i$th state symbol $S_{i}$ as $X_{i}=\textrm{\ensuremath{f_{x}}}(u_{1i}(M_{1},M_{2}),u_{2i}(M_{2}),S_{i})$.
Decoder 2 decodes the codeword $u_{2}^{n}(m_{2})$, while decoder
1 decodes both codewords $u_{2}^{n}(m_{2})$ and $u_{1}^{n}(m_{1},m_{2})$.
Using standard arguments, the rates (\ref{eqn: act_stat_DBC}) are
easily shown to be achievable. The proof of the converse can be found
in Appendix B.
\begin{rem}
If we let $p(s|a)=p(s)$ so that the channel is not action-dependent,
Proposition 6 recovers Proposition 4 of \cite{Steinberg_IT05} (see
also \cite{Kim}). 
\end{rem}

\subsection{A Binary Example}

In this section, we consider a special case of the model in Fig. \ref{fig:fig2}
in which the action channel $p(s|a)$ is binary and given by
\begin{eqnarray}
S & = & A\oplus B,\label{eq:act_chan}
\end{eqnarray}
where the action $A$ is binary, $B\sim\textrm{Ber}(b)$ and the transmission
channels are given by\begin{subequations}\label{eqn: DBC}
\begin{eqnarray}
Y_{1} & = & X\oplus S\oplus Z_{1},\label{eq:bin_dec1}\\
\textrm{and }Y_{2} & = & Y_{1}\oplus\widetilde{Z}_{2},\label{eq:bin_dec2}
\end{eqnarray}
\end{subequations}where $Z_{1}\sim\textrm{Ber}(N_{1})$ and $\widetilde{Z}_{2}\sim\textrm{Ber}(\widetilde{N}_{2})$
are independent of each other and of $B$. We select the cost metric
as $\gamma(a,x)=x$. We define $N_{2}=N_{1}*\widetilde{N}_{2}=N_{1}(1-\widetilde{N}_{2})+\widetilde{N}_{2}(1-N_{1})$. 

As a first remark, consider the ideal system with $b=0$ (i.e., no
interference) and no cost constraint (i.e., $\Gamma=\nicefrac{1}{2}$).
The system reduces to a standard physical degraded binary symmetric
broadcast channel, and thus the capacity region is given by the union
over $\alpha\in[0,0.5]$ of the rate pairs satisfying the inequalities
\cite[p. 115]{Elgammal}\begin{subequations}\label{eqn: DBC_bin}
\begin{eqnarray}
R_{1} & \leq & H(\alpha*N_{1})-H(N_{1})\label{eq:R1_BC_bin}\\
\textrm{and }R_{2} & \leq & 1-H(\alpha*N_{2}).\label{eq:R2_BC_bin}
\end{eqnarray}
\end{subequations}

We observe that, by construction, this rate region sets an outer bound
on the rate achievable in the system at hand. The outer bound above
is in fact achievable by setting $X=B$, $U_{2}\sim\textrm{Ber}(\nicefrac{1}{2})$,
$U_{1}=U_{2}\oplus\widetilde{U}_{1}$ with $\widetilde{U}_{1}\sim\textrm{Ber}(\alpha)$,
and $A=U_{1}$ in (\ref{eqn: act_stat_DBC}), where $U_{2}$ and $\widetilde{U}_{1}$
are independent. This entails that, by leveraging the actions, the
interference-free capacity region (\ref{eqn: DBC_bin}) is obtained
for all cost constraints $\Gamma\geq b$. It can be instead seen that,
if one is forced to set $A$ to be constant, achieving the rate region
(\ref{eqn: DBC_bin}) requires a cost $\Gamma=\nicefrac{1}{2}$, since
$X$ needs to be distributed $\textrm{Ber}(\nicefrac{1}{2})$. This
example illustrates the advantage of being able to affect the state
via actions selected as a function of the messages.

\subsection{Probing Capacity of Degraded Broadcast Channels\label{sub:Probing}}

In this section, we apply the setting of probing capacity introduced
in \cite{probing} to the degraded broadcast channel. Following \cite{probing},
the state sequence $S^{n}$ is thus assumed to be generated i.i.d.
according to a pmf $p(s)$. Moreover, based on the messages $(M_{1},M_{2})$,
the encoder selects an action sequence as in (\ref{eq:action_enc_DBC}).
However, here, through the choice of actions, the encoder affects
the state \textit{information} available at the encoder and the decoders,
and not the state sequence $S^{n}$. Specifically, the encoder obtains
partial state information $S_{e,i}=\textrm{\ensuremath{b_{e}}}(S_{i},A_{i})$,
and the decoders obtain partial state informations $S_{d_{1,i}}=\textrm{\ensuremath{b_{d{}_{1}}}}(S_{i},A_{i})$
and $S_{d_{2,i}}=\textrm{\ensuremath{b_{d{}_{2}}}}(S_{d_{1,i}})$,
respectively, where $i\in[1,n]$, and $b_{e}:\mathcal{S}\times\mathcal{A}\rightarrow\mathcal{S}_{e}$,
$b_{d_{1}}:\mathcal{S}\times\mathcal{A}\rightarrow\mathcal{S}_{d_{1}}$
and $b_{d_{2}}:\mathcal{S}_{d_{1}}\rightarrow\mathcal{S}_{d_{2}}$
are deterministic functions for given alphabets $\mathcal{S}_{e}$,
$\mathcal{S}_{d_{1}}$ and $\mathcal{S}_{d_{2}}$. Note that the state
information available at decoder 2 is degraded with respect to that
of decoder 1 (i.e., it is a function of the latter). As in \cite{probing},
we assume that the state information at the encoder is characterized
as
\begin{eqnarray}
S_{e,i}=\textrm{\ensuremath{b_{e}}}(S_{i},A_{i}) & = & \begin{cases}
\begin{array}{ccccc}
S_{i} &  &  & \textrm{if} & A_{i}=1\\
* &  &  & \textrm{if} & A_{i}=0
\end{array} & ,\end{cases}\label{eq:logic}
\end{eqnarray}
where $*$ represents the absence of channel state information at
the encoder. Moreover, the state information $S_{e,i}$ is assumed
to be available causally at the encoder so that the encoding function
is $g_{2i}\textrm{: }\mathcal{M}_{1}\times\mathcal{M}_{2}\times\mathcal{S}_{e}^{i}\rightarrow\mathcal{X}$
(cf. (\ref{eq:enc_DBC})). The rest of the code definition is similar
to Definition \ref{def_DBC} with the caveat that the decoder 1 and
2 have available also the information sequences $S_{d_{1}}^{n}$ and
$S_{d_{2}}^{n},$ respectively.

We note that, given the definitions above, the distribution of the
random variables $(M_{1},M_{2},A^{n},$ $S^{n},X^{n},Y_{1}^{n},Y_{2}^{n})$
factorizes as
\begin{align}
p(m_{1},m_{2},a^{n},s^{n},s_{e}^{n},s_{d_{1}}^{n},s_{d_{2}}^{n},x^{n},y_{1}^{n},y_{2}^{n})=\frac{1}{2^{n(R_{1}+R_{2})}}\delta[a^{n}-\textrm{\ensuremath{g_{1}}}(m_{1},m_{2})]\nonumber \\
\cdot\left\{ \prod_{i=1}^{n}p(s_{i})\delta[s_{e,i}-\textrm{\ensuremath{b_{e}}}(s_{i},a_{i})]\delta[s_{d_{1,i}}-\textrm{\ensuremath{b_{d{}_{1}}}}(s_{i},a_{i})]\delta[x_{i}-\textrm{\ensuremath{g_{2}}}(m_{1},m_{2},s_{e}^{i})\right\} \nonumber \\
\cdot\left\{ \prod_{i=1}^{n}\delta[s_{d_{2,i}}-\textrm{\ensuremath{b_{d{}_{2}}}}(s_{d_{1,i}})]p(y_{1i},y_{2i}|x_{i},s_{i},a_{i})\right\} \label{eq:code_dist_DBC_prob}
\end{align}
where the arguments of the pmf range in the alphabets of the corresponding
random variables. 

As discussed below, the setting at hand, which we refer to as having
a probing encoder, is a special case of the one studied in Sec. \ref{sub:System-model_act}.
Therefore, we can leverage Proposition \ref{prop:act_stat_DBC} to
obtain the following result.
\begin{prop}
The capacity region of the system in Fig. \ref{fig:fig2} under the
degradedness condition (\ref{eq:degradedness}) and with a probing
encoder is given by the union of the rate pairs $(R_{1},R_{2})$ satisfying
\begin{subequations}\label{eqn: act_stat_DBC_prob}
\begin{eqnarray}
R_{1} & \leq & I(U_{1};Y_{1},S_{d_{1}}|U_{2})\label{eq:R1_BC_prob}\\
\textrm{and }R_{2} & \leq & I(U_{2};Y_{2},S_{d_{2}})\label{eq:R2_BC_prob}
\end{eqnarray}
\end{subequations}where the mutual informations are evaluated with
respect to the joint pmf
\begin{eqnarray}
p(a,u_{1},u_{2},s,s_{e},s_{d_{1}},s_{d_{2}},x,y_{1},y_{2}) & = & p(s)p(u_{1},u_{2})\delta[a-\textrm{\ensuremath{f_{a}}}(u_{1},u_{2})]\delta[s_{e}-\textrm{\ensuremath{b_{e}}}(s,a)]\label{eq:joint_DBC_prob}\\
 &  & \cdot\delta[s_{d_{1}}-\textrm{\ensuremath{b_{d{}_{1}}}}(s,a)]\delta[s_{d_{2}}-\textrm{\ensuremath{b_{d{}_{2}}}}(s_{d_{1}})]\nonumber \\
 &  & \cdot\delta[x-\textrm{\ensuremath{f_{x}}}(u_{1},u_{2},s_{e})]p(y_{1}|x,s,a)p(y_{2}|y_{1}),\nonumber 
\end{eqnarray}
for some pmf $p(u_{1},u_{2})$ and deterministic functions $\textrm{\ensuremath{f_{a}}}\textrm{: }\mathcal{U}_{1}\times\mathcal{U}_{2}\rightarrow\mathcal{A}$
and \textup{$\textrm{\ensuremath{f_{x}}}\textrm{: }\mathcal{U}_{1}\times\mathcal{U}_{2}\times\mathcal{S}_{e}\rightarrow\mathcal{X}$}
such that the inequality\textup{ $\textrm{E}\left[\gamma(A,X)\right]\leq\Gamma$}
is satisfied.\end{prop}
\begin{IEEEproof}
The result is obtained by noticing that the setting described above
is a special case of the one described in Sec. \ref{sub:System-model_act_DBC}
by making the following substitutions\begin{subequations}\label{eqn: act_stat_DBC_prob_sub}
\begin{eqnarray}
S & \rightarrow & S_{e}\label{eq:R1_BC_prob-1}\\
\textrm{and }Y_{j} & \rightarrow & (Y_{j},S_{d_{j}})\textrm{ for }j=1,2.\label{eq:R2_BC_prob-1}
\end{eqnarray}
\end{subequations}To see this, we show that the pmf (\ref{eq:code_dist_DBC_prob})
reduces to (\ref{eq:code_dist_DBC}) under the given substitutions.
Specifically, by marginalizing (\ref{eq:code_dist_DBC_prob}) over
$S^{n}$ we have
\begin{eqnarray}
\frac{1}{2^{n(R_{1}+R_{2})}}\delta[a^{n}-\textrm{\ensuremath{g_{1}}}(m_{1},m_{2})]\left\{ \prod_{i=1}^{n}\delta[x_{i}-\textrm{\ensuremath{g_{2}}}(m_{1},m_{2},s_{e}^{i})]\delta[s_{d_{2,i}}-\textrm{\ensuremath{b_{d{}_{2}}}}(s_{d_{1,i}})]p(y_{2i}|y_{1i})\right\} \nonumber \\
\prod_{i=1}^{n}\overset{}{\underset{s_{i}\in\mathcal{S}}{\sum}}\left\{ p(s_{i})\delta[s_{e,i}-\textrm{\ensuremath{b_{e}}}(s_{i},a_{i})]\delta[s_{d_{1,i}}-\textrm{\ensuremath{b_{d{}_{1}}}}(s_{i},a_{i})]p(y_{1i}|x_{i},s_{i},a_{i})\right\} .\label{eq:before_cat}
\end{eqnarray}
The terms outside the summation in (\ref{eq:before_cat}) are equal
to the corresponding terms in (\ref{eq:code_dist_DBC}) under the
substitutions (\ref{eqn: act_stat_DBC_prob_sub}). For the remaining
terms, we observe that, for $a_{i}=0$, we have $S_{e,i}=*$, and
thus $p(s_{e,i}|a_{i})=\delta[s_{e,i}-*]$ and
\begin{eqnarray*}
p(y_{1i},s_{d_{1i}}|x_{i},s_{e,i},a_{i}) & = & \overset{}{\underset{s_{i}\in\mathcal{S}}{\sum}}p(s_{i})\delta[s_{d_{1,i}}-\textrm{\ensuremath{b_{d{}_{1}}}}(s_{i},a_{i})]p(y_{1i}|x_{i},s_{i},a_{i});
\end{eqnarray*}
instead, for $a_{i}=1$ we have $S_{e,i}=S_{i}$, and thus $p(s_{e,i}|a_{i})=\textrm{Pr}[S_{i}=s_{e,i}]$
and 
\begin{eqnarray*}
p(y_{1i},s_{d_{1i}}|x_{i},s_{e,i},a_{i}) & = & \delta[s_{d_{1,i}}-\textrm{\ensuremath{b_{d{}_{1}}}}(s_{e,i},a_{i})]p(y_{1i}|x_{i},s_{e,i},a_{i}),
\end{eqnarray*}
which completes the proof. 
\end{IEEEproof}

\section*{Concluding Remarks}

Action-dependent channels are useful abstractions of two-phase communication
scenarios. This paper has reported on two variations on this theme,
namely the problem of message and state transmission in an action-dependent
channel and the degraded action-dependent broadcast channel. Under
given assumptions, we have characterized the information-theoretic
performance of these systems. The analytical results, and specific
examples, emphasize the importance of jointly designing the transmission
strategy across the two communication phases.

\appendices{}

\section*{Appendix A: Proof of Proposition \ref{prop:act_stat}}

We first observe that given the probability of error constraint (\ref{eq:small_err})
we have the Fano inequality 
\begin{eqnarray}
H(M|Y^{n}) & \leq & n\delta(\epsilon),\label{eq:Fano_mess}
\end{eqnarray}
where the notation $\delta(\epsilon)$ represents any function such
that $\delta(\epsilon)\rightarrow0$ as $\epsilon\rightarrow0$, and
that given the CR constraint (\ref{eq:CK_req}), we have the Fano
inequality\begin{subequations}\label{eqn: Fano}
\begin{eqnarray}
H(\psi|Y^{n})\leq n\delta(\epsilon).\label{eq:Fano}
\end{eqnarray}
\end{subequations}We can then write
\begin{eqnarray}
nR & = & H(M)\overset{(a)}{\leq}I(M;Y^{n})+n\delta(\epsilon)\\
 & \overset{(b)}{=} & I(M;Y^{n})-I(M;S^{n}|A^{n})+n\delta(\epsilon)\\
 & = & I(\psi,M;Y^{n})-I(\psi;Y^{n}|M)-I(\psi M;S^{n}|A^{n})+I(\psi;S^{n}|A^{n},M)+n\delta(\epsilon)\\
 & = & I(\psi,M;Y^{n})-H(\psi|M)+H(\psi|M,Y^{n})-I(\psi,M;S^{n}|A^{n})+H(\psi|A^{n},M)\\
 &  & -H(\psi|A^{n},M,S^{n})+n\delta(\epsilon)\nonumber \\
 & = & I(\psi,M;Y^{n})\negmedspace-\negmedspace I(\psi;A^{n}|M)\negmedspace+\negmedspace H(\psi|M,Y^{n})\negmedspace-\negmedspace I(\psi,M;S^{n}|A^{n})\negmedspace\\
 &  & -\negmedspace H(\psi|A^{n},M,S^{n})\negmedspace+\negmedspace n\delta(\epsilon)\nonumber \\
 & \overset{(c)}{\leq} & I(\psi,M;Y^{n})-I(\psi,M;S^{n}|A^{n})+n\delta(\epsilon)\\
 & \overset{}{=} & \sum_{i=1}^{n}I(\psi,M;Y_{i}|Y^{i-1})-I(\psi,M;S_{i}|S_{i+1}^{n},A^{n})+n\delta(\epsilon)\\
 & \overset{(d)}{\leq} & \sum_{i=1}^{n}H(Y_{i})-H(Y_{i}|Y^{i-1},\psi,M,S_{i+1}^{n},A^{n})-H(S_{i}|S_{i+1}^{n},A^{n})\\
 &  & +H(S_{i}|Y^{i-1},\psi,M,S_{i+1}^{n},A^{n})+n\delta(\epsilon)\nonumber \\
 & \overset{(e)}{=} & \sum_{i=1}^{n}H(Y_{i})-H(Y_{i}|U_{i})-H(S_{i}|A_{i})+H(S_{i}|U_{i},A_{i})+n\delta(\epsilon)\\
 & \overset{}{=} & \sum_{i=1}^{n}I(U_{i};Y_{i})-I(U_{i};S_{i}|A_{i})+n\delta(\epsilon)\label{eq:conv_end}
\end{eqnarray}
where ($a$) follows due to Fano's inequality as in (\ref{eq:Fano_mess});
($b$) follows using the Markov chain $M-A^{n}-S^{n}$; (\emph{c})
follows by (\ref{eq:Fano}) and since mutual information is non-negative
(recall that by definition $2n\delta(\epsilon)=n\delta(\epsilon)$);
($d$) follows using the same steps provided in the proof of Theorem
1 in \cite[eq. (9)-(12)]{Weissman} by substituting $M$ with $(M,\psi)$
; and ($e$) follows by defining $U_{i}\overset{\bigtriangleup}{=}(Y^{i-1},\psi,M,S_{i+1}^{n},A^{n\backslash i})$
and because we have the Markov relation $S_{i}-A_{i}-(S_{i+1}^{n},A^{n\backslash i})$.

Defining $Q$ to be a random variable uniformly distributed over $[1,n]$
and independent of $(A^{n},S^{n},U^{n},X^{n},Y^{n})$, and with $A\overset{\triangle}{=}A_{Q}$,
$S\overset{\triangle}{=}S_{Q}$, $X\overset{\triangle}{=}X_{Q}$,
$Y\overset{\triangle}{=}Y_{Q}$ and $U\overset{\triangle}{=}(U_{Q},Q),$
from (\ref{eq:conv_end}) we have
\begin{eqnarray}
R & \leq & I(U;Y|Q)-I(U;S|A,Q)+\delta(\epsilon)\nonumber \\
 & \overset{(a)}{=} & H(Y|Q)-H(Y|U)-H(S|A,Q)+H(S|A,U)+\delta(\epsilon)\nonumber \\
 & \overset{(b)}{\leq} & H(Y)-H(Y|U)-H(S|A)+H(S|A,U)+\delta(\epsilon)\nonumber \\
 & = & I(U;Y)-I(U;S|A)+\delta(\epsilon)
\end{eqnarray}
where ($a$) follows using the definition of $U$ and ($b$) follows
because conditioning reduces entropy. Moreover, from (\ref{eq:cost_const}),
we have
\begin{eqnarray}
\Gamma+\epsilon & \geq & \frac{1}{n}\sum_{i=1}^{n}\textrm{E}\left[\gamma(A_{i},X_{i})\right]=\textrm{E}\left[\gamma(A,X)\right].
\end{eqnarray}
Next, define $\hat{S}_{i}=\psi_{i}\textrm{(}S^{n})$ and $\hat{S}=\hat{S}_{Q}$,
where $\psi_{i}\textrm{(}S^{n})$ represents the $i$th symbol of
$\psi\textrm{(}S^{n})$. Moreover, let $\mathcal{B}$ be the event
$\mathcal{B}=\{\psi\textrm{(}S^{n})\neq h_{2}(Y^{n})\}$. Using the
CR requirement (\ref{eq:CK_req}), we have $\textrm{Pr}(\mathcal{B})\leq\epsilon$.
We can then calculate the distortion as (we drop the dependence of
$\textrm{h}_{2i}$ on $Y^{n}$ for simplicity of notation) 
\begin{eqnarray}
\textrm{E}\left[d(S,\hat{S})\right]=\frac{1}{n}\sum_{i=1}^{n}\textrm{E}\left[d(S_{i},\hat{S}_{i})\right] & = & \frac{1}{n}\sum_{i=1}^{n}\textrm{E}\left[d(S_{i},\hat{S}_{i})\Big|\mathcal{B}\right]\textrm{Pr}(\mathcal{B})+\frac{1}{n}\sum_{i=1}^{n}\textrm{E}\left[d(S_{i},\hat{S}_{i})\Big|\mathcal{B}^{c}\right]\textrm{Pr}(\mathcal{B}^{c})\nonumber \\
 & \overset{(a)}{\leq} & \frac{1}{n}\sum_{i=1}^{n}\textrm{E}\left[d(S_{i},\hat{S}_{i})\Big|\mathcal{B}^{c}\right]\textrm{Pr}(\mathcal{B}^{c})+\epsilon D_{max}\nonumber \\
 & \overset{(b)}{\leq} & \frac{1}{n}\sum_{i=1}^{n}\textrm{E}\left[d(S_{i},h_{2i})\right]+\epsilon D_{max}\nonumber \\
 & \overset{(c)}{\leq} & D+\epsilon D_{max},
\end{eqnarray}
where ($a$) follows using the fact that $\textrm{Pr}(\mathcal{B})\leq\epsilon$
and that the distortion is upper bounded by $D_{max}$; ($b$) follows
by the definition of $\hat{S}_{i}$ and $\mathcal{B}$; and ($c$)
follows by (\ref{eq:dist_const}). 

To bound the cardinality of auxiliary random variable $U$, we first
observe that the distribution of the variables $(A,U,S,X,Y,\hat{S})$
identified above factorizes as
\begin{align}
p(a,u,s,x,y)=p(u)p(a,s,x|u)p(y|x,s,a) & ,\label{eq:joint_card}
\end{align}
and $\hat{S}$ is a deterministic function $\phi(U)$. Therefore,
for fixed $p(y|x,s,a)$, the characterization in Proposition \ref{prop:act_stat}
can be expressed in terms of integrals $\int g_{j}(p(a,s,x|u))dF(u)$
for $j=1,...,|\mathcal{A}|\times|\mathcal{S}|\times|\mathcal{X}|+2$,
of functions $g_{j}(.)$ that are continuous over pmf on the alphabet
$|\mathcal{A}|\times|\mathcal{S}|\times|\mathcal{X}|$. Specifically,
we have $g_{j}$ for $j=1,...,|\mathcal{A}|\times|\mathcal{S}|\times|\mathcal{X}|-1$
given by $p(a,s,x)$ for all values of $a\in\mathcal{A}$, $s\in\mathcal{S}$,
and $x\in\mathcal{X}$ (except one); $g_{|\mathcal{A}|\times|\mathcal{S}|\times|\mathcal{X}|}=H(Y|U=u)$;
$g_{|\mathcal{A}|\times|\mathcal{S}|\times|\mathcal{X}|+2}=H(S|A,U=u);$
and $g_{|\mathcal{A}|\times|\mathcal{S}|\times|\mathcal{X}|+1}=\textrm{E}\left[d(S,\hat{S})|U=u\right]$.
The cardinality bound follows by invoking Fenchel-Eggleston-Caratheodory
Theorem \cite[Appendix C]{Elgammal}. We finally observe that the
joint distribution (\ref{eq:joint_card}) can be written as (\ref{eq:joint})
without loss of generality, since $U$ can always contain $A$ without
reducing rate (\ref{eq:C}).

\section*{Appendix B: Proof of Proposition \ref{prop:act_stat_DBC}}

The proof is similar to that given in \cite{Steinberg_IT05} (see
also \cite{Kim}), although care must be taken to properly account
for the presence of the actions. We first observe that given the probability
of error constraint (\ref{eq:small_err_DBC}), we have the Fano inequality
$H(M_{j}|Y_{j}^{n})\leq n\delta(\epsilon)$ for $j=1,2.$ We can then
write the inequalities
\begin{eqnarray}
nR_{2} & = & H(M_{2})\overset{(a)}{\leq}I(M_{2};Y_{2}^{n})+n\delta(\epsilon)\\
 & \overset{(b)}{=} & \sum_{i=1}^{n}I(M_{2};Y_{2i}|Y_{2}^{i-1})+n\delta(\epsilon)\\
 & \overset{(c)}{\leq} & \sum_{i=1}^{n}I(M_{2},Y_{2}^{i-1};Y_{2i})+n\delta(\epsilon)\\
 & \overset{(d)}{\leq} & \sum_{i=1}^{n}I(M_{2},Y_{2}^{i-1},Y_{1}^{i-1};Y_{2i})+n\delta(\epsilon)\\
 & \overset{(e)}{=} & \sum_{i=1}^{n}I(U_{2i};Y_{2i})+n\delta(\epsilon),\label{eq:conv_end-2}
\end{eqnarray}
where ($a$) follows due to Fano's inequality as in (\ref{eq:Fano_mess});
($b$) follows by using the chain rule for mutual information; ($c$)
and ($d$) follow because conditioning increases entropy; and ($e$)
follows by defining $U_{2i}\overset{\bigtriangleup}{=}(M_{2},Y_{1}^{i-1})$
and noting the Markov relation $Y_{2}^{i-1}-(Y_{1}^{i-1},M_{2})-Y_{2i}$
due to the degradedness property (\ref{eq:degradedness}). We also
have the inequalities
\begin{eqnarray}
nR_{1} & = & H(M_{1})\overset{(a)}{\leq}I(M_{1};Y_{1}^{n})+n\delta(\epsilon)\\
 & \overset{(b)}{\leq} & I(M_{1};Y_{1}^{n}|M_{2})+n\delta(\epsilon)\\
 & \overset{(c)}{=} & \sum_{i=1}^{n}I(M_{1};Y_{1i}|Y_{1}^{i-1},M_{2})+n\delta(\epsilon)\\
 & \overset{(d)}{\leq} & \sum_{i=1}^{n}I(M_{1},Y_{1}^{i-1},S^{i-1};Y_{1i}|Y_{1}^{i-1},M_{2})+n\delta(\epsilon)\\
 & \overset{(d)}{\leq} & \sum_{i=1}^{n}I(U_{1i};Y_{1i}|U_{2i})+n\delta(\epsilon),\label{eq:conv_end-2-1}
\end{eqnarray}
where ($a$) follows due to Fano's inequality as in (\ref{eq:Fano_mess});
($b$) follows because $M_{1}$ and $M_{2}$ are independent and since
conditioning reduces entropy; ($c$) follows using the chain rule
for mutual information; ($d$) follows since conditioning decreases
entropy; and ($e$) follows by defining $U_{1i}\overset{\bigtriangleup}{=}(M_{1},Y_{1}^{i-1},S^{i-1})$.
Let $Q$ be a random variable uniformly distributed over $[1,n]$
and independent of $(A^{n},S^{n},U_{1}^{n},U_{2}^{n},X^{n},Y_{1}^{n},Y_{2}^{n})$
and define $A\overset{\triangle}{=}A_{Q}$, $S\overset{\triangle}{=}S_{Q}$,
$X\overset{\triangle}{=}X_{Q}$, $Y_{1}\overset{\triangle}{=}Y_{1Q}$,
$Y_{2}\overset{\triangle}{=}Y_{2Q}$, $U_{1}\overset{\triangle}{=}(U_{1Q},Q)$,
and $U_{2}\overset{\triangle}{=}(U_{2Q},Q)$. We easily see that,
with these definitions, the sum (\ref{eq:conv_end-2}) is upper bounded
by $I(U_{2};Y_{2})$, and (\ref{eq:conv_end-2-1}) equals $I(U_{1};Y_{1}|U_{2})$.
Moreover, note that, from the definitions above, $X$ is a function
of $U_{1}$, $U_{2}$ and $S$, given the encoding function (\ref{eq:enc_DBC}).
Similarly, $A$ is a function of $(U_{1},U_{2})$ given (\ref{eq:action_enc_DBC}).
We also have the Markov relationship $(U_{1},U_{2})-A-S$ as it can
be easily checked by using the d-separation principle \cite{Kramer}.
Finally, from (\ref{eq:cost_const}), we have
\begin{eqnarray*}
\Gamma+\epsilon & \geq & \frac{1}{n}\sum_{i=1}^{n}\textrm{E}\left[\gamma(A_{i},X_{i})\right]=\textrm{E}\left[\gamma(A,X)\right],
\end{eqnarray*}
which completes the proof.

\section*{Acknowledgment}

The work of O. Simeone is supported by the U.S. National Science Foundation
under grant CCF-0914899.

\ifCLASSOPTIONcaptionsoff \newpage{}\fi

\end{document}